\documentclass[11pt,twoside]{article}

\usepackage{asp2006}
\usepackage{epsf}
\usepackage{psfig}
\usepackage{lscape}

\begin{document}
\title{Only the Lonely: HI Imaging of Void Galaxies}   
\author{K. Stanonik\altaffilmark{1}, E. Platen\altaffilmark{2}, M. A. Arag\'on-Calvo\altaffilmark{3}, J. H. van Gorkom\altaffilmark{1}, R. van de Weygaert\altaffilmark{2}, J. M. van der Hulst\altaffilmark{2},  K. Kova\v{c}\altaffilmark{4}, C.-W. Yip\altaffilmark{3}, P. J. E. Peebles\altaffilmark{5}}   
\altaffiltext{1}{Department of Astronomy, Columbia University, 550 West 120th Street, New York, NY 10027, USA; email: kstanonik@astro.columbia.edu}
\altaffiltext{2}{Kapteyn Astronomical Institute, University of Groningen, PO Box 800, 9700 AV Groningen, The Netherlands}
\altaffiltext{3}{The Johns Hopkins University, 3701 San Martin Drive, Baltimore, MD 21218, USA}
\altaffiltext{4}{Institute of Astronomy, ETH Z\"{u}rich, CH-8093, Z\"{u}rich, Switzerland}
\altaffiltext{5}{Joseph Henry Laboratories, Princeton University, Princeton, NJ 08544, USA}

\begin{abstract} 
We have completed a pilot survey imaging 15 SDSS selected void galaxies in H \textsc{i} in local (d=50 to 100 Mpc) voids. 
This small sample makes up a surprisingly interesting collection of galaxies, consisting of galaxies with asymmetric and perturbed H \textsc{i} disks, 
previously unidentified companions, and ongoing interactions. One was found to have a polar H \textsc{i} disk with no 
stellar counterpart. While our small number statistics so far are limiting, results support past findings that most 
void galaxies are typically late type galaxies with gas rich disks and small scale clustering similar to field galaxies 
despite their large scale underdense environment.
\end{abstract}

\section{Project Description}

\begin{figure}[!ht] 
\plotone{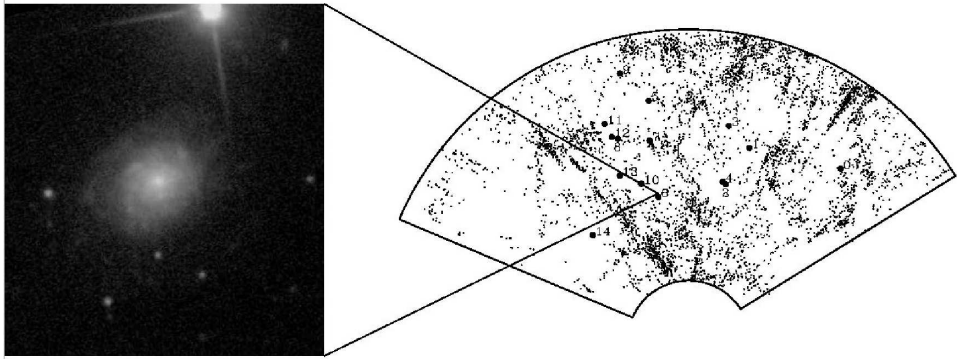}
\caption{Location of void galaxies within the SDSS redshift survey.
\label{fig:sdssloc}}
\end{figure}

The advent of wide and deep redshift surveys has allowed the definitive identification of void galaxies by their 3D environmental underdensity.  We have selected a sample of 60 void galaxies from the Sloan Digital Sky Survey (SDSS) using a novel, purely structural and geometric technique  (see talk by R. van de Weygaert and E. Platen).  Galaxies were chosen to be centrally located within the deepest underdensities, as reconstructed at a scale of 1 Mpc/h from the SDSS galaxy redshift survey distribution at distances of 50-100 Mpc (Fig. \ref{fig:sdssloc}).

Targets were observed in full 12 hour tracks at the Westerbork Synthesis Radio Telescope (WSRT) with a typical velocity resolution of 8.6 km/s, spatial resolution of 15$^{\prime\prime}$, and rms of 0.4 mJy/beam.  Typical WSRT sensitivity gives a 3$\sigma$ detection limit of $\sim 2 \times 10^8 M_{\sun}$ for a galaxy with velocity width of 100 km/s at a distance of 70 Mpc, and a column density sensitivity of $\sim 5 \times 10^{19} $cm$^{-2}$.  We use a powerful new backend at the WSRT which probes a velocity range of nearly 10,000 km/s, allowing the detection of faint nearby companions which may be interacting, as well as background galaxies which provide a higher density control sample also contained within the SDSS volume.  

\section{Pilot Sample Results}

\begin{figure}[!hb] 
\plotfiddle{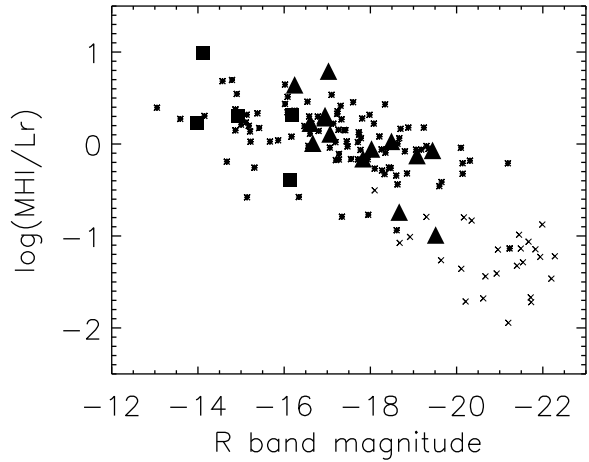}{2.5in}{90.}{35.}{35.}{150}{0}
\caption{H \textsc{i} mass to light ratio.  Triangles are our target galaxies and squares are the companion galaxies. +Õs are taken from \cite{Swaters2002} and XÕs are taken from  \cite{Verheijen2001}.   
\label{fig:mhilr}}
\end{figure}

We have examined an initial pilot sample of 15 void galaxies (Stanonik et al. in prep). Their H \textsc{i} masses range from $3.5 \times10^8$ to $3.8 \times 10^9$ M$_\odot$, with one non-detection with a 3$\sigma$ upper limit of $2.1 \times 10^8 $ M$_\odot$ assuming a velocity width of 150 km/s. 
We found most of our void galaxies exhibit disk-like rotation and contain an extended supply of gas. Many have very blue optical colors and large H$\alpha$ equivalent widths, suggesting high rates of recent star formation.  We find that their H \textsc{i} mass is typical for their size, with our smallest galaxies less efficient at turning gas into stars (Fig. \ref{fig:mhilr}).

\clearpage

\begin{figure}[!ht]
\plottwo{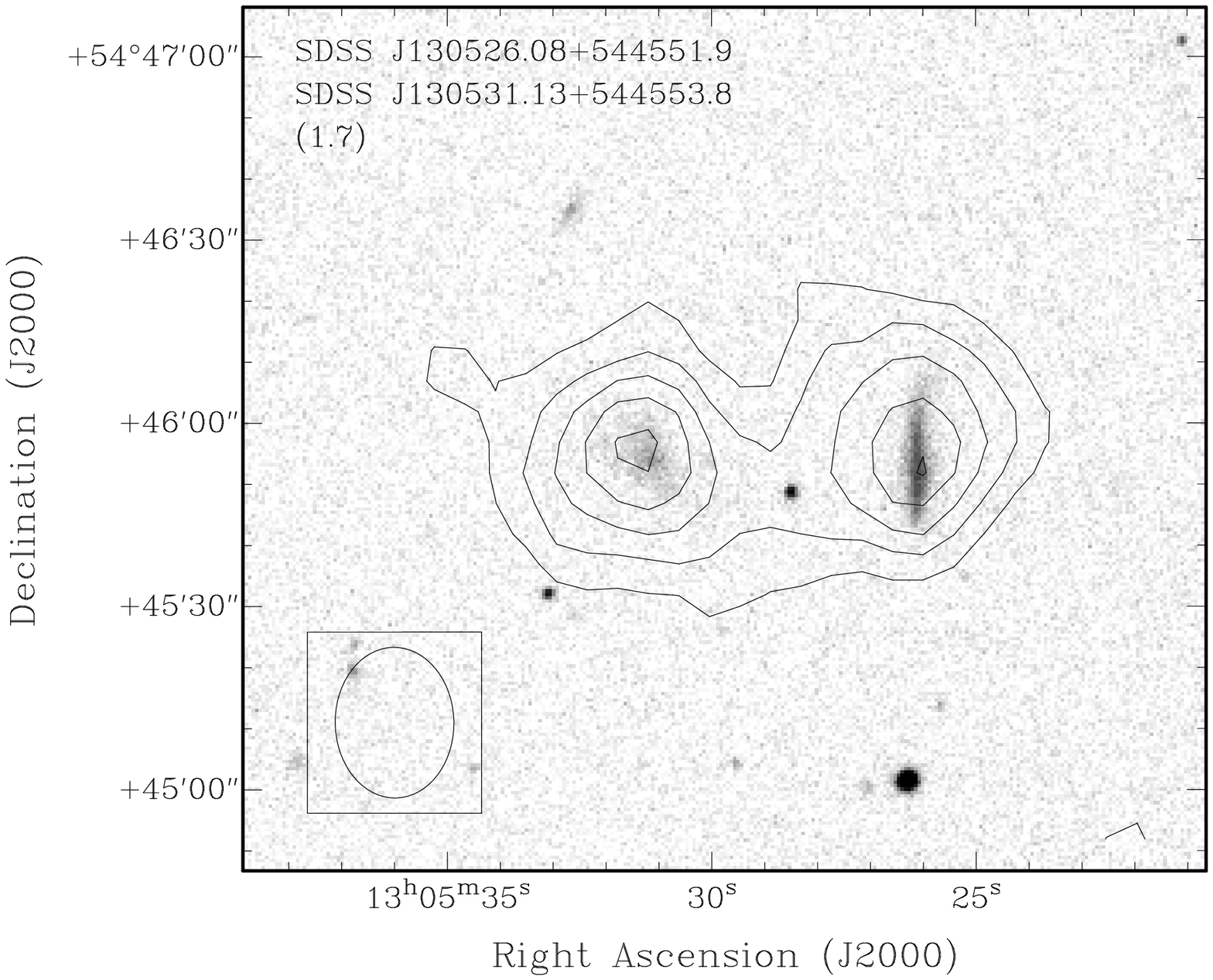}{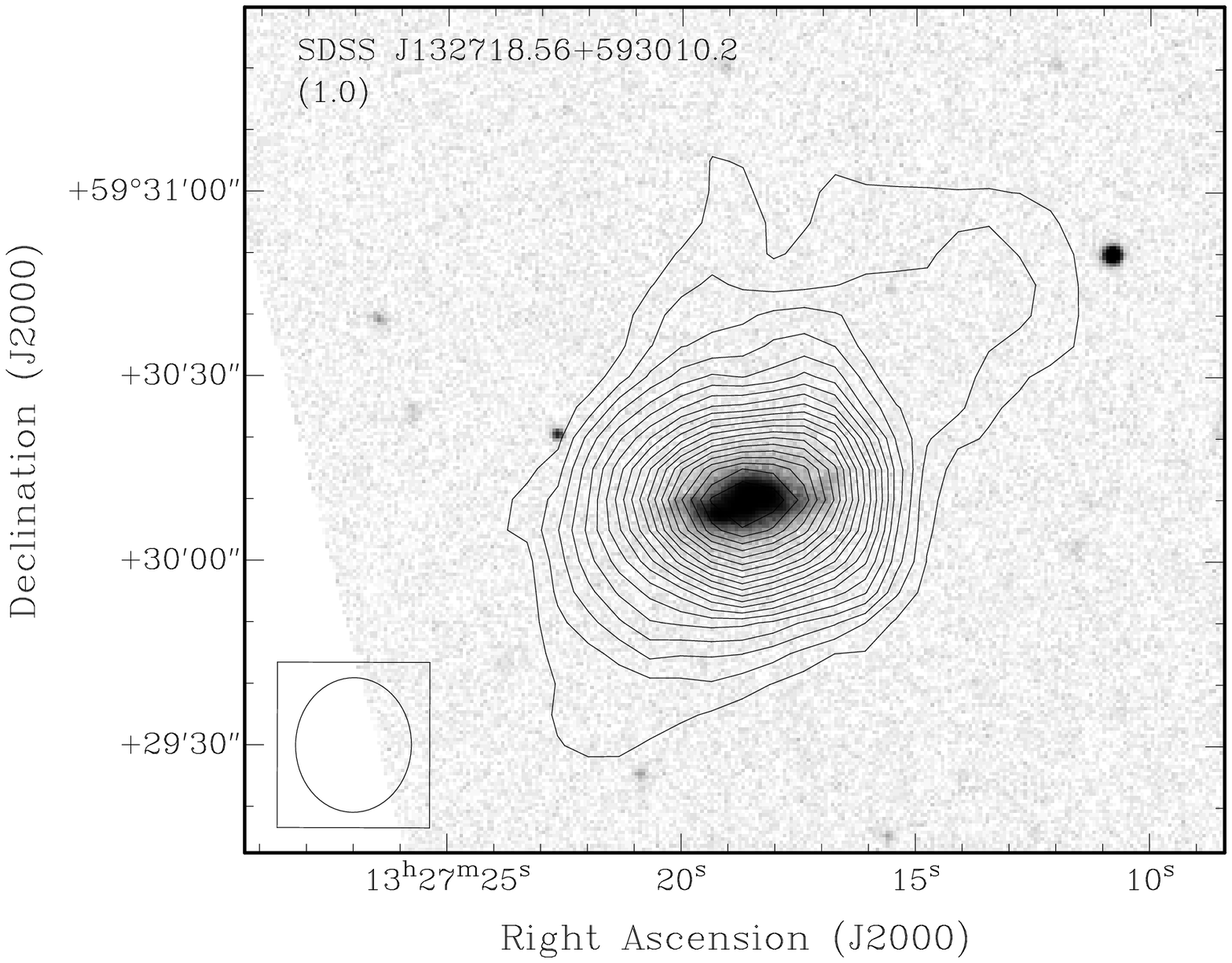}
\plottwo{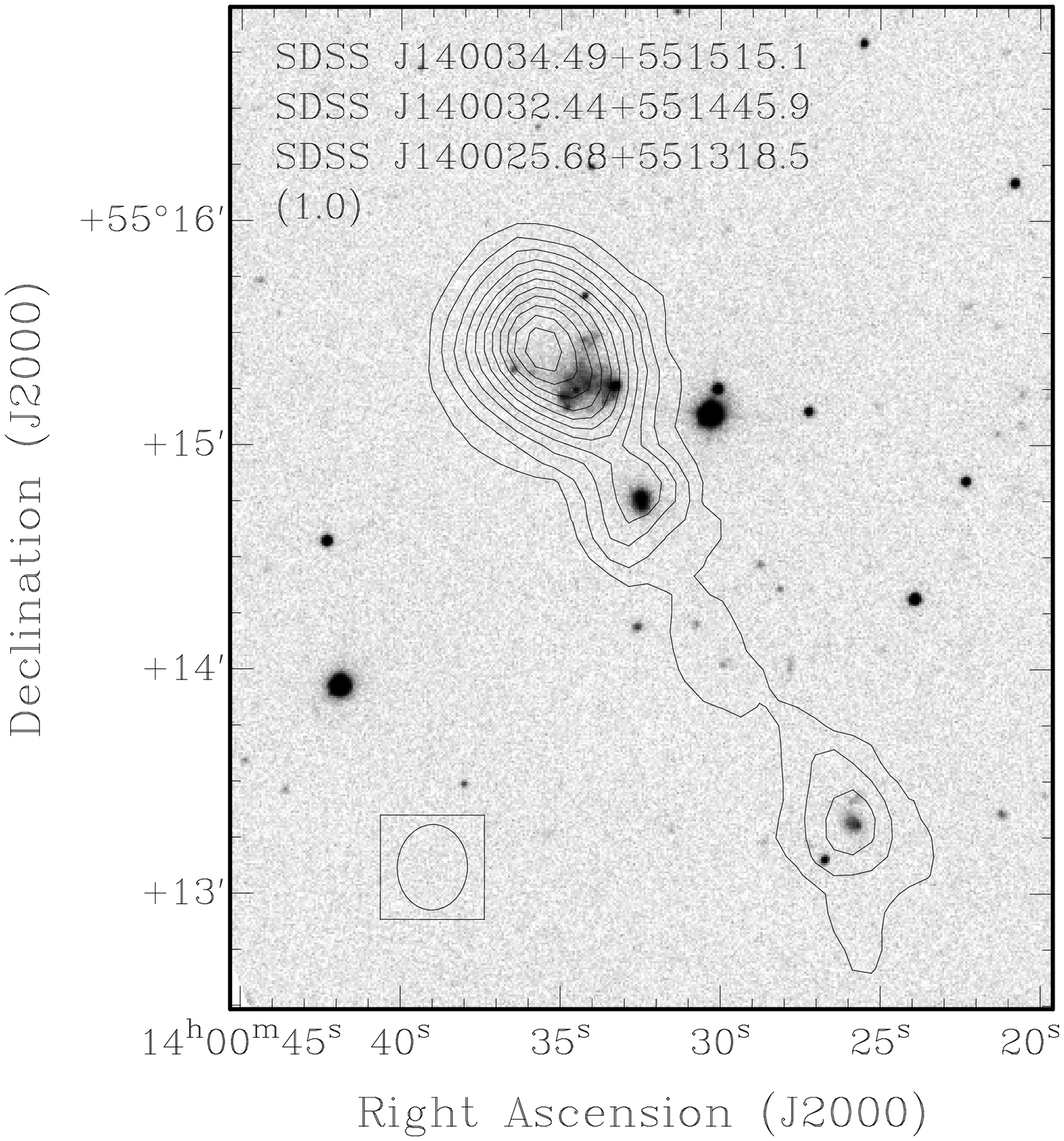}{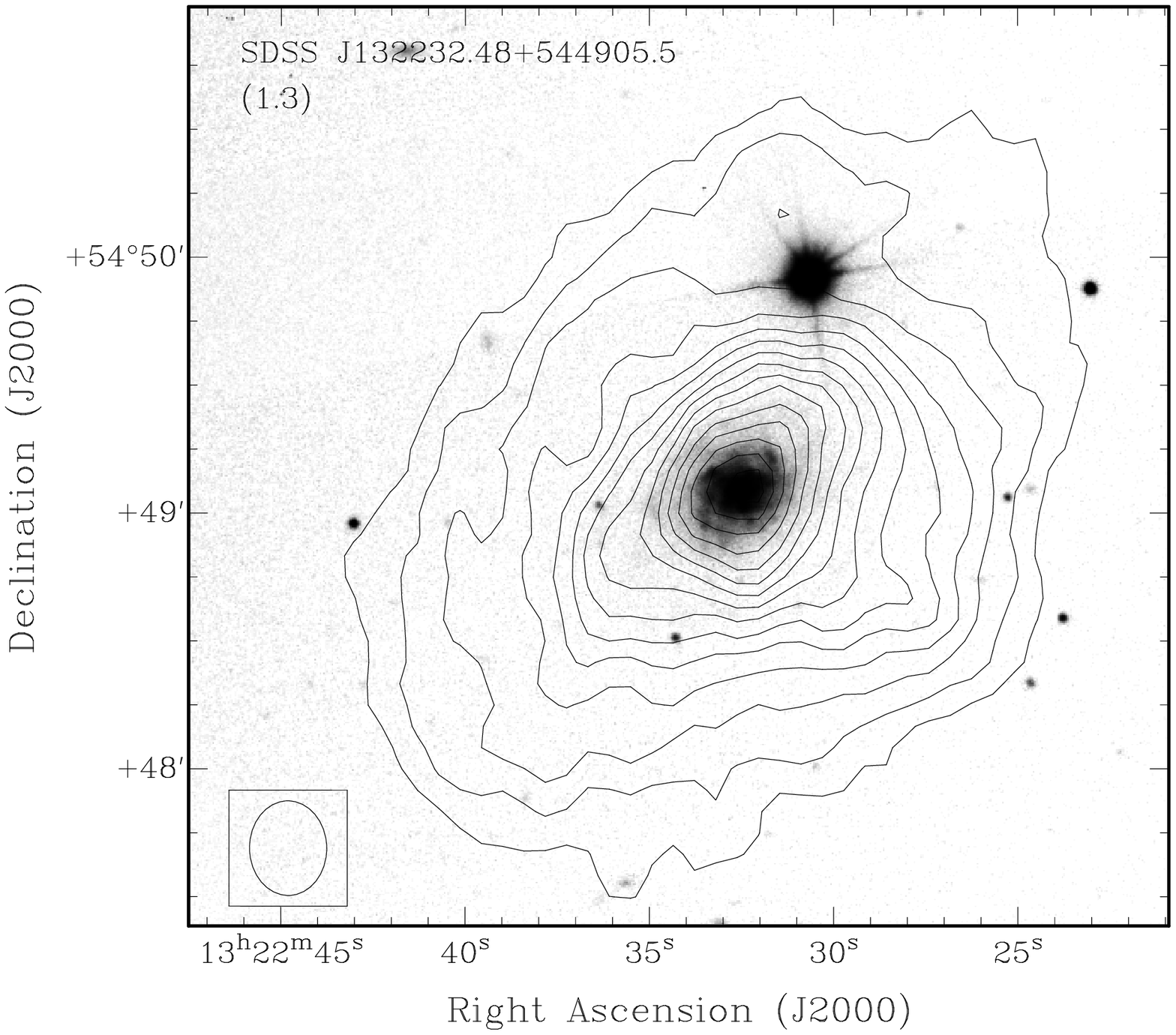}
\caption{Lonely but not alone.  SDSS g-band images are overlaid with contours from the total intensity maps at $5 \times 10^{19} cm^{-2}$ plus increments of $10^{20} cm^{-2}$ for four of our targeted void galaxies. Confidence level ($\sigma$) of the lowest contour is given in the top left corner of each image.  On the left, galaxies interacting with H \textsc{i} rich companions.  On the right, galaxies asymmetric in their H \textsc{i} distribution.
\label{fig:pics}}
\end{figure}

Many void galaxies were found to appear fairly irregular optically or to have asymmetries in their H \textsc{i} distribution at the lowest contour levels (Fig. \ref{fig:pics}, right).   
In addition, some galaxies were found to have faint companions (Fig. \ref{fig:pics}, left), supporting the findings of \cite{Szomoru1996} and \cite{Abbas2007} that the small scale clustering of galaxies in voids is unaffected by the large scale underdensity.  In our sample of 15 galaxies, five were found to have H \textsc{i} rich companions, two of which are interacting in H \textsc{i} with the targeted void galaxy. In these cosmologically defined void regions, the large scale ($\sim$20 Mpc) underdensity does not require local ($\sim$1 Mpc) isolation, though it is generally the case.
We hope our complete sample will more statistically address this issue and perhaps provide some insight into the question of the ``missing'' dwarf galaxy population predicted by $\Lambda$CDM simulations \citep{Peebles2001}.  

\clearpage

\begin{figure}[!ht] 
\plottwo{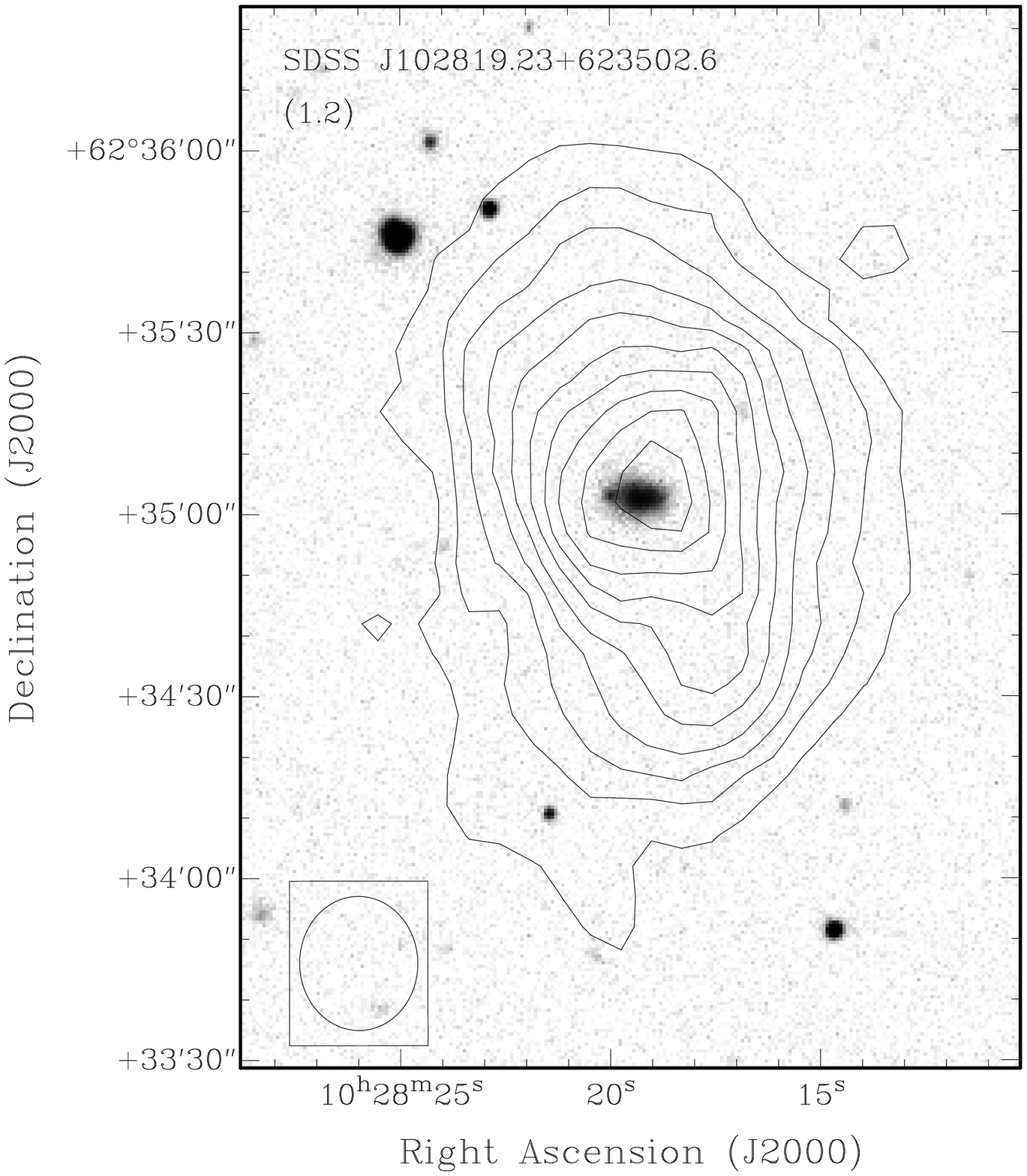}{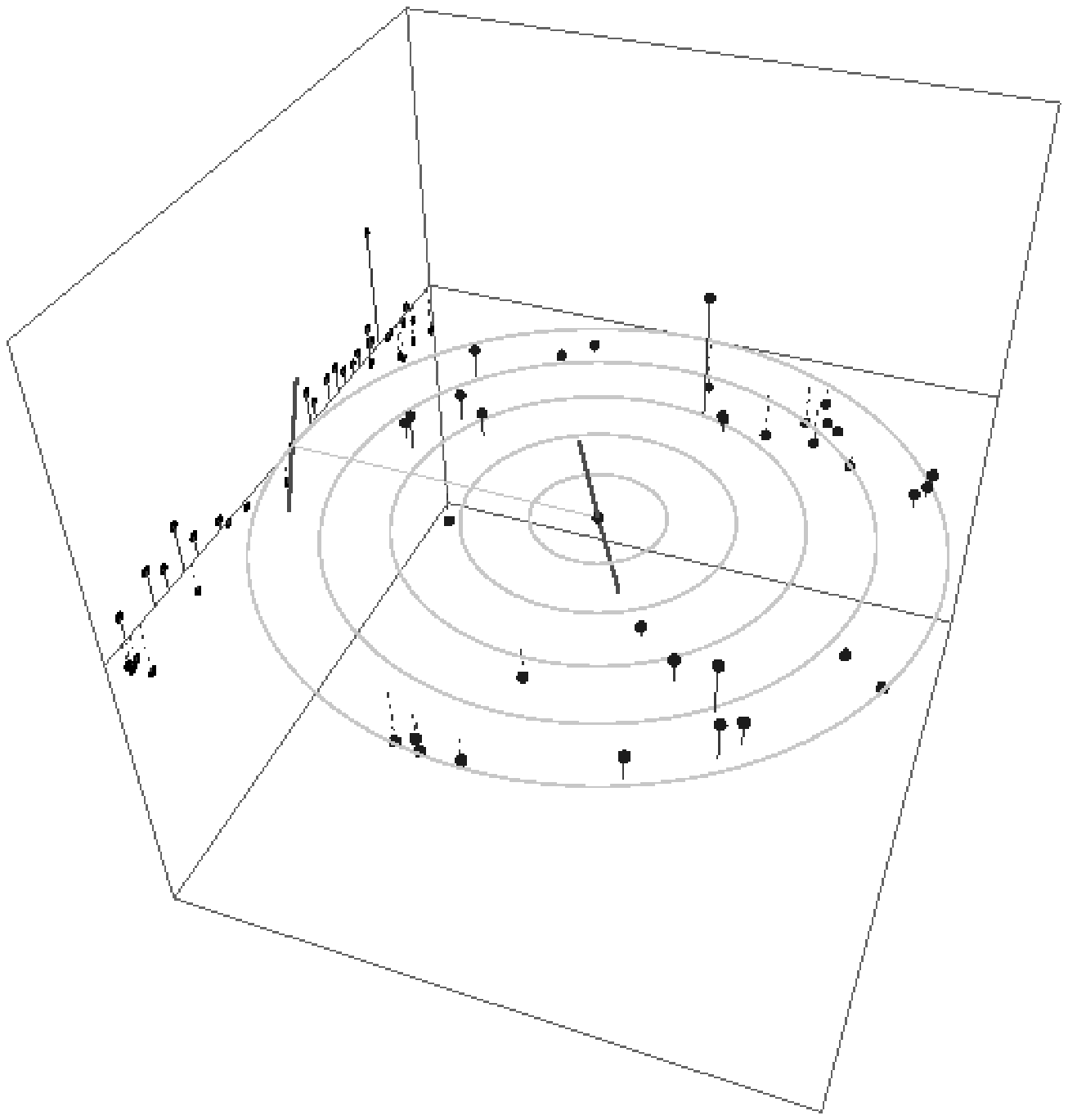}
\caption{On the left, the SDSS g-band image, overlaid with $5 \times 10^{19} cm^{-2} + 10^{20} cm^{-2}$ H \textsc{i} contours.  On the right, the location and orientation in the SDSS of the polar disk within a thin wall, between two voids. Concentric circles mark every 2 Mpc in the plane of the wall.  An edge-on view is projected on the left, showing the thinness of the wall. The red line indicates the position and orientation of the projected major axis of the H \textsc{i} disk. 
\label{fig:pdg} }
\end{figure}

These galaxies are interesting both statistically and individually. One was found to have a polar disk in H \textsc{i} with no optical counterpart in the SDSS or Medium Sky Survey NUV images (Fig. \ref{fig:pdg}), yet the H \textsc{i} mass in the disk ($3 \times 10^9 M_\odot$) is comparable to the stellar mass in the galaxy. This suggests slow accretion of the H \textsc{i} material at a relatively recent time by cold mode accretion (Stanonik et al. 2009).  This galaxy is located within a diffuse wall in the SDSS, between two voids.  The polar disk is oriented nearly perpendicular to the wall, implying that this significant mass of H \textsc{i} has been accreted out of the voids.  

We hope that our full sample will allow a more careful analysis of the gas and star formation properties of void galaxies as a detailed function of their environment.  In addition to presenting a well constrained test of $\Lambda$CDM predictions, this population may provide key insights into many other aspects of galaxy formation and evolution.

\acknowledgements 
This work was supported in part by the National Science Foundation under grant \#0607643 to Columbia University.

\end{document}